\documentclass{emulateapj}
\shorttitle{The Once and Future Andromeda Stream}
\shortauthors{Mori and Rich}

\begin{document}

\title{The Once and Future Andromeda Stream}

\author{Masao Mori\altaffilmark{1}}
\affil{Department of Law, Senshu University, Higashimita 2-1-1, 
Tama, Kawasaki 214-8580, Japan}
\email{mmori@isc.senshu-u.ac.jp}
\and
\author{R. Michael Rich}
\affil{Department of Physics and Astronomy, University of California, 
Los Angeles, CA 90095-1562, USA}
\email{rmr@astro.ucla.edu}

\altaffiltext{1}{Visiting researcher of Department of Physics and Astronomy, 
University of California, Los Angeles} 

\begin{abstract}
The interaction between an accreting satellite and the Andromeda galaxy 
(M31) has been studied analytically and numerically, using a high-resolution 
$N$-body simulation with $4\times10^7$ particles. For the first time, we show 
the self-gravitating response of the disk, the bulge, and the dark matter 
halo of M31 to an accreting satellite. We reproduce the Stream and the shells 
at the East and West side of M31, by following the evolution of the collision 
4 Gyr into the future, and show that recently discovered diffuse arcs on the 
south side of the minor axis are the remnants of a similar collision that 
occurred 3-4 Gyr earlier than the Stream event. The present day integrity of 
the M31 disk constrains the satellite mass to be $M \leq 5\times10^9 M_\odot$. 
The stars that were originally in the center of the satellite are now in the 
east shell. Therefore, observations in this region might reveal additional 
clues about the nature of satellite, such as the central core and any 
metallicity gradient. 
\end{abstract}


\section{Introduction}
Large spiral galaxies like the Andromeda galaxy (M31) are believed to have
formed in part from the merger of many less massive galaxies, and their halos 
are modeled as the debris of accreted dwarfs (e.g. Bullock \& Johnston 2005).
The giant stream of Ibata et al. (2001) and the starcount maps of Ferguson 
et al. (2002) establish the complex merger origin of the M31 halo population; 
evidence for coherent structures now extends to $\approx 100 $ kpc (Ibata et 
al. 2007). Further support comes from the spectroscopic side e.g. the disk 
population of Ibata et al. (2005). Deep HST/ACS photometry shows that three 
deep fields in the M31 halo exhibit unequivocal evidence for intermediate age 
populations (Brown et al. 2006). Starcount maps and radial velocities of red 
giant stars near M31 exhibit a giant stellar stream to the south of this 
galaxy, as well as giant stellar shells to the east and the west of M31's 
center (Ibata et al. 2001; Ferguson et al. 2002; McConnachie et al. 2003; 
Ibata et al. 2004, 2005; Guhathakurta et al. 2006). Recently Ibata et al. 
(2007) have carried out a deep, wide field photometric survey of the minor 
axis toward M33. They find a series of arc-like structures parallel to the 
major axis of M31 toward M33, and spaced by $\approx 10$ kpc.

$N$-body simulations of the interaction between the progenitor of the giant 
stream and M31 suggest that the stream and shells are the tidal debris 
formed in the last pericentric passage of a satellite on a radial orbit 
(Ibata et al. 2004; Font et al. 2006; Geehan et al. 2006; Fardal et al. 2006, 
2007). However, these models have always assumed a fixed potential to 
represent the influence of M31, and no models consider the possible effect 
of a live ($N$-body) disk. The discovery of the widespread intermediate 
age halo population has raised the question of whether these stars belong 
to the infalling satellite or are instead disk stars that were ejected in 
the stream collision or an earlier interaction. Present constraints on the 
progenitor mass are weak; consequently we look to the present day thickness 
of the disk to limit the progenitor mass. In addition, the newly discovered 
large scale arcs (Ibata et al. 2007) have not yet been modeled.  These issues 
motivate us to explore the effect of the accreting satellite using the first 
self-consistent, $N$-body model of M31 that has a disk, bulge, and dark matter 
halo.

\section{Disk Thickness constrains the Satellite Mass}

It has long been known that infalling satellites are likely the major source 
of the observed thickness of disks, and that the integrity of disks constrain
the masses of such satellites (T\'{o}th \& Ostriker 1992; Quinn, Hernquist 
\& Fullagar 1993; Vel\'{a}zquez \& White 1999; Font et al. 2001; Hayashi \& 
Chiba 2006; Gauthier, Dubinski \& Widrow 2006). 

We estimate the disk heating by dynamical friction, which is caused by the 
scattering of the disk stars into an overdense wake that trails the orbiting 
body, exerting a force opposite to the orbital motion. The energy input into 
disk stars through the interaction between disk stars and a satellite is equal 
to the orbital energy loss $\Delta E$ of the satellite. If we assume that the 
vertical velocity dispersion of the disk stars $\sigma_{\rm z}$ is constant, 
the change $\Delta \sigma_{\rm z}$ is simply denoted by 
$\Delta \sigma_{\rm z}^2 = 2 \Delta E / M_{\rm s}$, where $M_{\rm s}$ is the mass 
of the satellite, and $\Delta E$ is given by the integral of the frictional 
force over an orbit of the satellite within the disk. Using the (Chandrasekhar 
1943) expression for dynamical friction, for a thin disk, $\Delta E$ is  
$\Delta E \simeq 4 \pi ~G^2 M_{\rm s}^2  \Sigma / v_{\rm s}^2$, where $G$ is the 
gravitational constant, $\Sigma$ is the surface mass density of a disk and 
$v_{\rm s}$ is the satellite velocity.

For an axisymmetric thin disk, the equation of motion in the vertical 
direction and the Poisson equation are simply expressed as 
$\partial_{\rm z} (\rho_{\rm d} \sigma_{\rm z}^2)+\rho_{\rm d} \partial_{\rm z} \Phi=0$ 
and $\partial_{\rm z}^2 \Phi=4 \pi G \rho_{\rm d}$, where $\rho_{\rm d}$ is the 
mass density of the disk and $\Phi$ is the gravitational potential, 
respectively. Using these equations and assuming the 
$\rho_{\rm d} \propto \exp(-z/z_{\rm d})$, where $z_{\rm d}$ is the scale height 
of the disk, we obtain the relation between $\sigma_{\rm z}$ and $z_{\rm d}$: 
$\sigma_{\rm z}^2=2 \pi G \Sigma z_{\rm d}$. 
Inserting the equation of $\Delta \sigma_{\rm s}^2$ into the change of the 
vertical velocity dispersion of the disk as $\Delta \sigma_{\rm z}^2=2 \pi G 
\Sigma \Delta z_{\rm d}$, we get the relation as 
$M_{\rm s} = [ M_{\rm d} v_{\rm s}^2 \Delta z_{\rm d}/(4 \, G) ]^{1/2}$.
Using a set of parameters given by Model A of Widrow, Perrett \& Suyu 
(2003), finally, we obtain the critical satellite mass, 

\begin{eqnarray}
M_{\rm c} &=& 5.2 \times 10^9 M_\odot \nonumber \\ 
&\times& \left( \frac{M_{\rm d}}{7 \times 10^{10} ~M_\odot} \right)^{1/2}
   \left( \frac{v_{\rm s}}{150 ~{\rm km ~s}^{-1}} \right)
   \left( \frac{\Delta z_{\rm d}}{0.3 ~{\rm kpc}} \right)^{1/2}.
\end{eqnarray}

Consequently, the dynamical mass of the satellite should be smaller than 
this critical mass, since the disk thickness must agree with the observed 
thickness of M31 after the interaction of the satellite. {\it This yields 
an upper limit for the total progenitor mass.}  Adopting for the stream 
[Fe/H]$\ga -1$ (Koch et al. 2007) and using the mass-metallicity relation 
of Dekel \& Woo (2003) gives a lower mass limit for a progenitor stellar 
mass $\approx 5\times 10^8 M_\odot$, which agrees with the predictions of 
Font et al. (2006, 2007).
We offer only order of magnitude constraints on these mass bounds, which must 
be refined further with better observations and fully nonlinear numerical 
simulations.

%
% Fig. 1
%
\begin{figure*}
\begin{center}
\includegraphics[scale=1.]{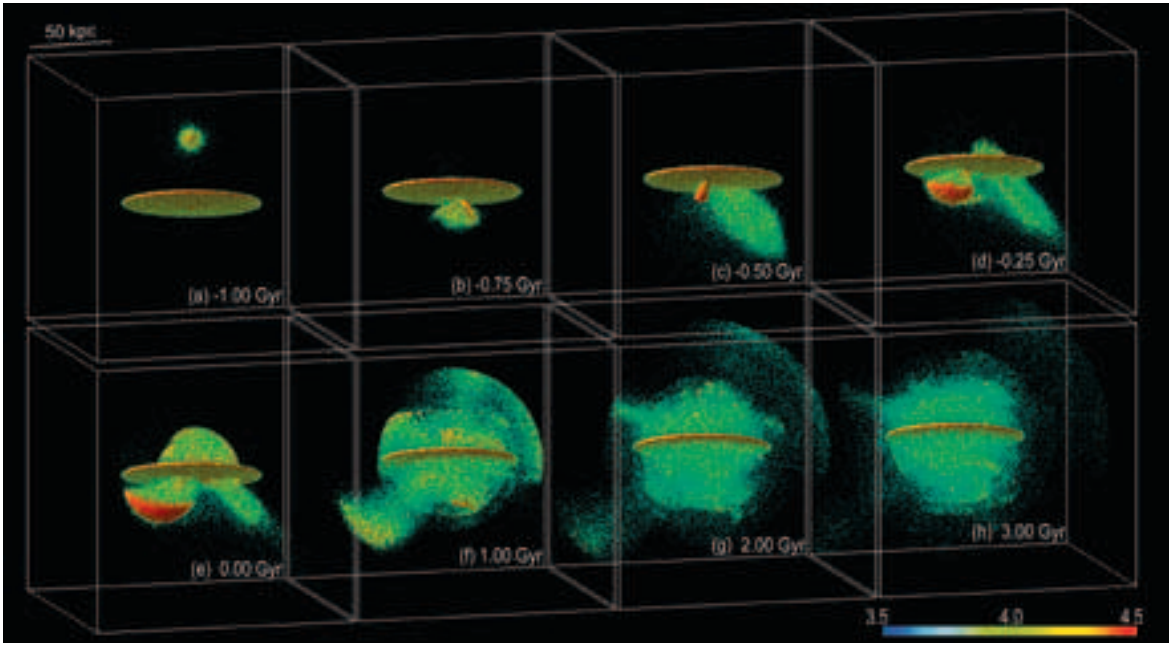}
%\epsscale{1.0}
%\plotone{f1.eps}
\caption{
Morphological evolution of the satellite. (a)-(f) The spatial distribution 
of stellar mass density as a function of the elapsed time, where 0.00 Gyr 
corresponds to the present. Each simulation box has a size of 160 kpc. 
The color corresponds to the logarithmic density $3.5 \leq \log \rho 
~(M_\odot {\rm ~kpc}^{-3}) \leq 4.5$. Elapsed time in Gyr is given at the 
lower-right corner in each panel: (a)$-1.00$ Gyr, (b)$-0.75$ Gyr, (c)$-0.50$ 
Gyr, (d)$-0.25$ Gyr, (e) 0.00 Gyr, (f) 1.00 Gyr, (g) 2.00 Gyr, and (h) 3.00 
Gyr, respectively. The formation of shells in the panel (e)-(h) infers that 
the distant arcs observed by Ibata et al. (2007) may correspond to the shells 
of an ancient radial merger. This figure is also available as an mpeg 
animation in the electronic edition of the {\it Astrophysical Journal}.
}
\label{f1}
\end{center}
\end{figure*}
%
%
% Fig. 2
%
\begin{figure}
\begin{center}
\includegraphics[scale=0.5]{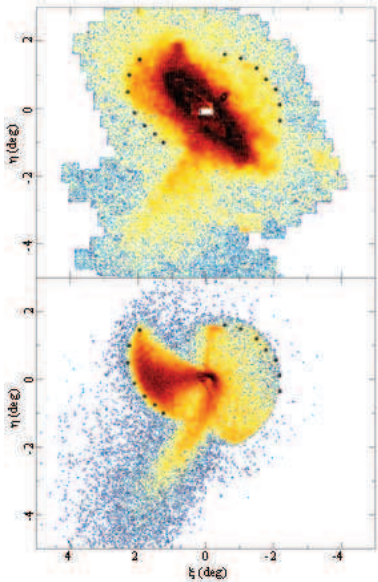}
%\epsscale{0.8}
%\plotone{f2.eps}
\caption{
Comparison between our simulation and the observed starcounts. $Upper$: 
The map of RGB count density around M31 observed by Irwin et al. (2005). 
$Lower$: The projected stellar density of satellite particles for Model A.
}
\label{f2}
\end{center}
\end{figure}
%
%
% Fig. 3
%
\begin{figure}
\begin{center}
\includegraphics[scale=0.5]{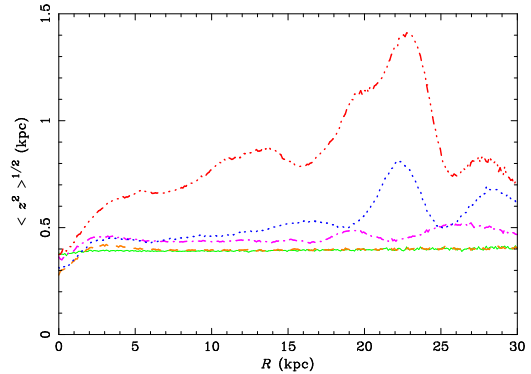}
%\epsscale{1.0}
%\plotone{f3.eps}
\caption{
Vertical extent of the disk at the present-day for different impact satellite 
masses. The solid line and the dashed line show the initial radial profile 
and the resultant radial profile after 5 Gyr without a collision, respectively.
The dot-dashed line, the dotted line and the dot-dot-dot-dashed line 
correspond to the satellite mass of $10^9 M_\odot$ (Model A), 
$5\times10^9 M_\odot$ (Model B) and $10^{10} M_\odot$ (Model C), respectively. 
We notice that our preferred Model A has a negligible effect on the disk.}
\label{f3}
\end{center}
\end{figure}

\section{Numerical modeling of the interaction between the satellite and M31}

In this section, we demonstrate an $N$-body simulation of the interaction 
between an accreting satellite and M31. We assume that the total mass of the 
disk is $7.0\times10^{10} M_\odot$. The density of the disk falls off 
approximately as the exponential with the scale length of 5.4 kpc in the 
radial direction, and as sech$^2$ with the scale height of 0.3 kpc in the 
vertical direction. The bulge is a King (1966) model with the total mass of 
$2.5\times10^{10} M_\odot$. The dark matter halo is taken to be a lowered Evans 
model (Kuijken \& Dubinski 1995) with the total mass of 
$3.2\times10^{11} M_\odot$ and the tidal radius of 80 kpc. 
This set of parameters corresponds to Model A of Widrow, Perrett \& Suyu 
(2003), and provides a good match with the observational data for M31. An 
$N$-body realization of this model is done by {\sc GalactICS} code written 
by Kuijken \& Dubinski (1995). The satellite is assumed to be a Plummer 
sphere with initial mass $10^{9} M_\odot$ (Model A), $5\times10^9 M_\odot$ 
(Model B) and $10^{10} M_\odot$ (Model C). The initial scale radius is 1 kpc. 
Following Fardal et al. (2007), we adopt the initial position vector and 
velocity vector for the standard coordinates centered from M31 are 
$(-34.75, 19.37, -13.99)$ kpc and $(67.34, -26.12, 13.50)$ km s$^{-1}$, 
respectively. The evolution of the collisionless system is followed up to 
5 Gyr using the parallel $N$-body code {\sc Gadget-2} (Springel 2005) and 
{\sc AFD} (Mori \& Umemura 2006). We adopt a tree algorithm with a tolerance 
parameter of $\theta=0.5$ and a softening length is 50 pc for all particles. 
The number of particles is $7,351,266$ for the disk, $2,581,897$ for the bulge, 
and $30,750,488$ for the dark matter halo. For the satellite, we use $10^5$ 
particles for Model A, $5\times10^5$ particles for Model B, and $10^6$ 
particles for Model C. The total number of particles is $\approx 40$ million, 
and the mass of a particle is $10^4M_\odot$.

Figure 1 shows the results for the time sequence of stellar dynamics for 
Model A. The upper panels in double columns illustrate the spatial 
distribution of stellar density as a function of time from $-1$ to $-0.25$ 
Gyr (where 0 Gyr is present-day).  The lower panels describe the {\it future} 
evolution of the system from the present-day to 3 Gyr. In our time frame, 
$-1$ Gyr corresponds to the initial condition and the start of the simulation 
run. The first pericentric passage occurred about 0.8 Gyr ago. Figure 1b 
shows that the satellite collides almost head-on with the bulge. 
Then, the distribution of satellite particles is distorted and is spread out 
significantly as seen in Fig. 1c. A large fraction of the satellite particles 
acquires a high velocity relative to the center of M31. This debris expands 
at great distance, remaining collimated and it gives rise to the southern 
giant stream. The apocentric passage occurred $\sim 0.7$ Gyr ago and the 
second pericentric passage is shown in Fig. 1c. As seen in Fig. 1d, stellar 
particles that initially constituted the satellite start to form a clear 
shell structure after the second collision with the disk. Moreover, a double 
shell system is sharply defined in Fig. 1e. The system is composed of 
approximately constant curvature shells formed by phase wrapping (Hernquist 
\& Quinn 1988). Then, these shells further expand and a multiple large 
scale-shell system is finally formed in the outer region and the dense core 
is formed in the inner region. The outermost large-scale shells in our 
simulation have a radius of $ > 50$ kpc and these structures survive at least 
4 Gyr from the present-day.

Figure 2 shows the comparison between the map of RGB count density around 
M31 observed by Irwin et al. (2005) and the projected stellar density at the 
present-day in our simulation. The satellite is entirely disrupted, and the 
giant stream of debris arising from the tidal destruction of the accreting 
satellite at the southern part of M31 is observed. The total mass of the 
stream given by the simulation is $1.4\times10^8 M_\odot$. This is consistent 
with the estimated mass $\sim 3\times 10^8 M_\odot$ from the observations by 
Ibata et al (2001).
The maximum length and width of the present-day model stream are $\sim 150$ 
kpc and $\sim 50$ kpc, respectively, in good agreement with the observations 
of Ibata et al. (2007).
Furthermore, the simulation reproduces the butterfly-shaped shells in the 
northeast and the west part of M31. In the RGB count map of Irwin et al. 
(2005), the brightness of the northeast shell appears to exceed that of the 
west shell. The observed density contrast between the shells is 0.43 at 
$(\xi, \eta)=(2.0, 0.7)$ and $(-1.5, 0.7)$ with the radius of 0.5 deg for the 
standard coordinates centered from M31. On the other hand, the simulated 
density contrast is 0.25, which reasonably matches the observed faint 
features in M31.

Figure 3 shows the disk thickness $<z>^{1/2}$ as a function of the radius 
following the interaction, for different satellite models. The solid line 
and the dashed line show the initial radial profile and the resultant radial 
profile after 5 Gyr {\it without} the infalling satellite, respectively. 
A concordance of two lines confirms that our fiducial model of M31 has 
perfect dynamical stability during the integration time. There is no 
significant impact on the disk kinematics or disk thickness for Model A and B. 
Therefore, satellites less massive than $10^9 M_\odot$ have a negligible effect 
on the disk dynamics. But it is clear that the massive satellite (Model C) 
more effectively heats the disk than do the less massive satellites.
For the inner part of the disk, $R < 10$ kpc, the scale height increases less 
than $\sim 10\%$ of its initial value for Model A and B, while it increases 
more than $\sim 50\%$ for Model C. Thus, a massive satellite is ruled out as 
the progenitor of the giant stream, and this result confirms the discussion 
in \S 2.

\section{Discussion}

%
% Fig. 4
%
\begin{figure}
\begin{center}
\includegraphics[scale=0.8]{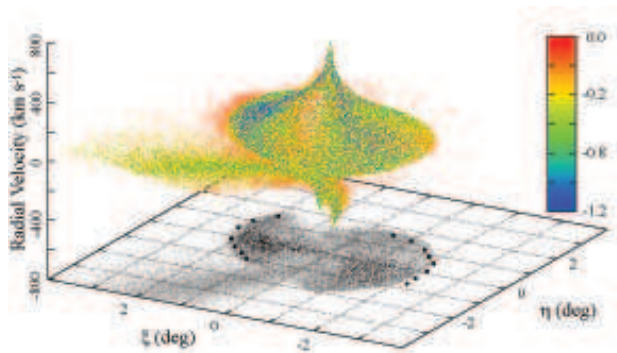}
%\epsscale{1.0}
%\plotone{f4.eps}
\caption{
Phase space distribution of the satellite particles. The color of each 
particle corresponds to the initial total energy with the unit of $10^{51}$ 
erg, with more negative values having originally been deeper in the potential. 
Note that most of the particles that were originally deep in the potential of 
the satellite are found at present day in the East shell.
}
\label{f4}
\end{center}
\end{figure}

In this work, we follow the approach of earlier studies (Ibata et al. 2004; 
Font et al. 2006; Geehan et al. 2006; Fardal et al. 2006, 2007) in simulating 
the infall of a massive satellite into M31. However, we add a live $N$-body 
disk and bulge, and this permits us to ask under what conditions this 
collision might eject stars from these populations into the halo and 
quantifies the effect of these populations on the satellite. We also follow 
the evolution of this encounter for 4 Gyr beyond the present.
As is the case in Fardal et al. (2007), our simulations successfully 
reproduce the giant stream and (for a similar but much older collision) the 
apparent shell structures that are observed in the starcount maps. 
The large number of test particles in our satellite affords a good comparison 
with observations (Koch et al. 2007). We conclude that the event responsible 
for the giant stream most likely has not populated the halo with ejected disk 
stars; hence the intermediate age metal rich populations observed in deep HST 
fields by Brown et al. (2006) are not disk stars ejected {\it by the stream 
progenitor.} An ancient radial collision involving a more massive progenitor 
($\approx 10^{10} M_\odot$), or a collision with a different approach angle, 
might have been capable of ejecting disk stars into the halo. The present-day 
thickness of the disk following this collision constrains the progenitor mass 
to be $\leq 5\times10^9 M_\odot$.

Our simulation gives additional predictions, such as the existence of high 
velocity stars in the central region of M31. Figure 4 shows that the tidal 
debris in the M31 central bulge has velocity $ > 300 {\rm ~km s}^{-1}$. 
The particles that were originally deepest in the potential (see color-coding 
in $10^{51}$ erg units) are likely, at present, to lie in the East shell. 
Observations in this region might reveal additional clues about the nature of 
satellite, such as the central core and the metallicity gradients (cf. Mori et 
al. 1997, 1999). 
It is also noteworthy that $>3$ Gyr after the collision, the debris more or 
less uniformly fills the halo.  Simulations such as Font et al. (2007) may 
wish to consider the disk, bulge, halo, and satellite stellar populations and 
their effect on the long term evolution of streams; encounters with stars in 
the disk and bulge may disrupt the coherence of streams and scatter the stars 
in a manner analogous to disk and bulge shocking of globular clusters.

A series of arc-like structures discovered by Ibata et al. (2007) are similar 
to the large-scale shell structures shown in Fig. 1f-1h. We conclude that 
these arcs may be the fossils of previous radial mergers several Gyr in the 
past (cf. Hernquist \& Quinn 1988).  Our model predicts that similar arcs 
should be found on the opposite side of the M31 disk, in the Northwest 
quadrant, if in fact they are the shells of ancient radial infall collisions.
Our model also offers a natural explanation for stars with $R>100$ kpc that 
are observed in the M31 halo near the systemic velocity of M31 (Gilbert et al. 
2006; Koch et al. 2007).  
Some of these stars at $R\approx 165$ kpc are found to have [Fe/H]=$-1$ by
Koch et al. (2007).  We propose that these stars are {\it not} a pressure 
supported halo but are rather the remnants of ancient radial collisions 
(note that in Fig. 1, the final extent of the collision is of order 100 kpc). 
Future studies will examine collisions from different attack angles, stellar 
disks, and the fate of dust and gas.

\acknowledgments
We are grateful to A. Koch, H. Ferguson, D. Reitzel, and
F. Schweizer for valuable discussions.  We thank M. J. Irwin for use of 
observational data, K. Shimasaku for data processing, and M. Umemura for 
use of the FIRST cluster. The computations were performed on the FIRST 
cluster at CCS, the University of Tsukuba and the SPACE at Senshu University. 
MM was supported in part by the Grant-in-Aid of the JSPS, 14740132, of the 
MEXT, 16002003, and of Senshu University, 2006, 2007.
RMR acknowledges support from GO-10265 and 10816 from the Space Telescope 
Science Institute, and grants AST-039731, and AST-0709479 from the National 
Science Foundation.

\end{document}